\documentclass[final, numberedheadings]{aipproc}

\usepackage{amsmath, amssymb, latexsym}

\layoutstyle{6x9}

\def \be{\begin{equation}}
\def \ee{\end{equation}}

\def \f+{\vec{f_{+}}}
\def \fx{\vec{f_{\times}}}

\begin{document}

\title{Optimal Source Tracking and Beaming of LISA}



\author{Archana Pai}{
address={Max-Planck Institut f\"ur Gravitationsphysik, Am M\"uhlenberg 1, 
14476 Potsdam, Germany},
}

\copyrightyear{2006}

\begin{abstract}
We revisit the directionally optimal data streams of LISA first
introduced in Nayak etal.  
It was shown that by using appropriate choice of Time delay
interferometric (TDI) combinations, a monochromatic fixed source in
the barycentric frame can be optimally tracked in the LISA frame.  In
this work, we study the beaming properties of these optimal
streams. We show that all the three streams $v_{+,\times,0}$ with
maximum, minimum and zero directional SNR respectively are highly beamed.  We study
in detail the frequency dependence of the beaming.
\end{abstract}

\date{\today}
\maketitle

\section{Introduction}
The space-based gravitational wave (GW) mission LISA \cite{LISArep} ---
Laser Interferometric Space Antenna --- consists of three identical
space-crafts forming an equilateral triangle of side {$5\times
10^{6}$} km following heliocentric orbit trailing the Earth by
{$20^\circ$}.  The plane of LISA makes an angle of {$60^\circ$} with
the plane of the ecliptic. Each space-craft completes one orbit around
the sun as well as in LISA plane in one year. The mission is aimed at
detecting and analyzing the low frequency GW signals in the frequency
range of $0.1 {\rm mHz} - 1 {\rm Hz}$. The astrophysical sources for
the LISA include galactic binaries, super-massive black-holes (BH),
extreme mass ratio inspirals, intermediate mass BHs.

Due to LISA's rotational as well as orbital motion, a fixed source in
the barycentric frame appears to follow a specific track in its sky.
This introduces amplitude modulation,
frequency modulation, 
and phase modulation 
in the 6 Doppler data streams.
A large number of interferometric configurations, having different
frequency and angular response, can be constructed from these data
streams which makes LISA not just a single detector but a network of
interferometers.

The choice of the combination of the data streams depend on the which
question one wishes to address. In Ref. \cite{NDPV03}, we addressed
the question of directional optimality in LISA i.e. for a given sky
location, which LISA data stream gives maximum SNR. It was shown that 
the
directional optimality condition 
gives three data streams: (1) {$v_+$} -- with maximum directional SNR,
(2) {$v_\times$} -- with minimum directional SNR, (3) {$v_0$} -- with
zero directional SNR. It was shown that the data stream $v_+$
optimally tracks the source motion (in the LISA sky) of a fixed source
in the barycentric frame. Here, {\it tracking} involves appropriate
choice of data combinations (switch combinations as source moves)
which gives maximum SNR in that direction. 
Tracking known monochromatic binaries with such streams
could give information about the source distance, polarizations.
For an unknown distant source, it would amount to 'looking' in
a specific direction in a particular frequency band. 

In this work, we study the beaming property of the optimal data streams
(in particular, $v_+$)
for a monochromatic source tracked for a year.
We show that $v_+,v_\times and v_0$ are beamed, i.e. they are sensitive
towards the tracking direction. 
The beam-width depends on the frequency under consideration.
We study the nature of this dependence.

The paper is organized as follows: In Sec \ref{sec:II}, we review the
TDI. In the first half of Sec.  \ref{sec:III}, we summarize the main
results of Ref.\cite{NDPV03} pertaining to the directionally optimal
TDI streams.  In the later half of Sec \ref{sec:III}, we discuss the
beaming properties of the directional streams followed by conclusion
in Sec. \ref{sec:IV}.


\section{TDI data streams}
\label{sec:II}
The 6 LISA Doppler data streams $W_\sigma^m$, where $m=1,2,3, \sigma
=\pm$, [see Fig.\ref{fig:lisa}{\footnote{$W_+^1=U^3, W_+^2=U^1,
W_+^3=U^2, W_-^1=-V^2, W_-^2=-V^3, W_-^3=-V^1$of \cite{TD04}}] are
obtained by letting the laser beams from each space-craft to travel
towards two other space-crafts and are beaten with the on board
laser. $m$ corresponds to the arm index and ($-$)$+$ indicates the
laser beam traveling in the (anti-)clockwise direction.

\begin{figure}[!htb]
\includegraphics[width=0.4\columnwidth,height=1.5in]{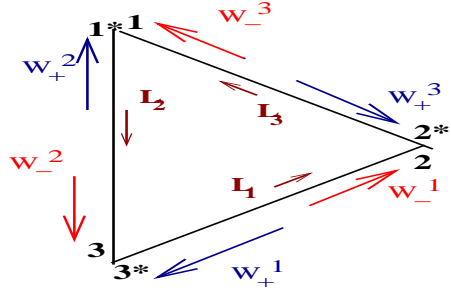}
\caption{The LISA constellation}
\label{fig:lisa}
\end{figure}

These Doppler data streams contain the phase fluctuation noise of the
master laser {($\Delta \nu /\nu_0 \simeq 10^{-13}/\sqrt{{Hz}}$)} which
is several orders of magnitude higher than the LISA designed sensitivity
level of {$h \sim 10^{-22}$}.  Using TDI technique, $W_\sigma^m$ can
be combined by appropriately delaying them with time delay operators
to construct laser noise-free data streams Ref. \cite{TD04}. 
The TDI scheme is based on the principle of sagnac interferometry ---
the light-beam is split and made to travel along two paths in opposite
direction of equal length which are then subtracted to remove the
laser noise. In TDI, equal path difference is
achieved by adding Doppler-streams with appropriate time delay operators
which are then added (or subtracted) to obtain laser noise free data stream.

Any laser noise-free data combination can be written as
${D = \sum_{m=1}^3 \sum_{\sigma=\pm} \rho_{m \sigma} W_{\sigma}^m}$,
where {$\rho_{m \sigma}$} are polynomials of time-delay operators
 {$E_m: E_m C_j (t) = C_j (t - L_m)$} and {$C_m$} are the
 laser phase noise fluctuation at the space-craft {$m$}.
The Sagnac TDI combinations {$\{\alpha,\beta,\gamma,\zeta\}$} in terms
of {$\{\rho_{m+},\rho_{m-}\}$} are given by \footnote{This set forms a
generator set for an algebraic module of all laser noise free data
combinations Ref. \cite{TD04}.}  {
\begin{eqnarray}
\alpha &=& \{E_2,1, E_1 E_2, -E_3, -E_1 E_3,-1 \} \hspace{1cm} \beta = \{E_2 E_3, E_3, 1,-1,-E_1,-E_1 E_2 \} \,,\\
\gamma &=& \{1, E_1 E_3, E_1, -E_2 E_3, -1, -E_2 \} \hspace{1cm} \zeta = \{E_3, E_1, E_2, -E_2, -E_3 , -E_1 \} \,.
\end{eqnarray}
}

The {$\zeta$} combination is termed as symmetrized sagnac due to its
symmetric structure and  is
insensitive to GW at low frequency $f << 1/L=60$ mHz ($L=16.7$ sec.). 

 The noise vector for any combination  {$D$} is given by
\begin{eqnarray}
N^{D} &=&( 2 \sqrt{S^{pf}} (\rho_{m+}^{D} + \mu_{m+}^{D}),2 \sqrt{S^{pf}} (\rho_{m-}^{D} + \mu_{m-}^{D}), \sqrt{S^{opt}} \rho_{m+}^{D}, \sqrt{S^{opt}} \rho_{m-}^{D})
\end{eqnarray}
where the polynomials {$\mu_{m\pm}$} are defined as {$\mu_{3-} = (E_3
\rho_{3+} - \rho_{3-})/2 = - \mu_{2+}$}. The rest $\mu_{m\sigma}$ can
be obtained by cyclic permutations. The {$S^{pf} = 2.5 \times 10^{-48}
(f/1 {\rm Hz})^{-2} {\rm Hz}^{-1}$} and the {$S^{opt} = 1.8 \times
10^{-37} (f/1 {\rm Hz})^2 {\rm Hz}^{-1}$} are the one-sided power
spectral densities (PSD) of the proof-mass noise and optical-path
noise respectively \cite{LISArep}. In the frequency domain
{$E_m=\exp(i \Omega L_m)$}.

For simplicity, throughout this work, we assume the three arms of LISA
to be equal i.e. {$L_m \equiv L$}. This helps in simplifying the
expressions of TDI streams and are exact for low frequencies. However
for higher frequencies, the above simplification leads to small
discrepancies.  Nevertheless, one can easily extend this for
unequal-arm interferometry.

A set of TDI data streams which diagonalizes the noise covariance
matrix  {$N^{(I)}\cdot N^*_{(J)}$} are \cite{PTL02}:
%
$Y^{(1)}=(\alpha+\beta-2 \gamma)/\sqrt{6}, 
Y^{(2)}=(\beta-\alpha)/\sqrt{2},
Y^{(3)}=(\alpha+\beta+\gamma)/\sqrt{3} \,.$
%
$Y^{(1)}$,
$Y^{(2)}$ and $Y^{(3)}$ are known as E,A and T in the LISA
literature.

\section{Directional Data Streams}
\label{sec:III}

As mentioned in the introduction, the three $Y^{(I)}$'s give different
frequency as well as angular response.  A large number of TDI data
streams {$\sum \alpha_{(I)} (f,\theta_L,\phi_L) Y^{(I)}$} with
different angular and frequency responses can be constructed from
them.  The choice of a combination depends on the question one wishes
to address.  In Ref. \cite{NDPV03}, we asked the following question:
If one wants to observe a particular sky location
$\{\theta_B,\phi_B\}$, in a given frequency bin, which {$\sum
\alpha_{(I)} Y^{(I)}$} TDI data stream would be optimal (maximum SNR)?
We first briefly summarize the results of Ref. \cite{NDPV03}.

The GW response of {$Y^{(I)}$} expressed in frequency domain is
{\small
\begin{equation}
h^{(I)}(\Omega)=F_{+}^{(I)}(\Omega)h_{+}(\Omega)+F_{\times}^{(I)}
(\Omega)h_{\times}(\Omega)\,,
\end{equation}
}
\hspace*{-0.3cm} 
where, the antenna pattern functions {$F_{+,\times}^{(I)}$} in terms
of the transfer function of each Doppler data stream is
{\small
\begin{equation} 
F_{+,\times}^{(I)}(\Omega) = i \sum_{m,\sigma}\rho_{m \sigma}^{(I)} \, 
\Delta \phi_m \, \exp(i \Omega (\hat{w} \cdot \hat{a}_m)) \, 
{\rm Sinc} (k_{m \sigma} \, \Delta \phi_m) \xi_{m ; +,\times} \, ,
\end{equation}
}
\vspace*{-0.5cm}

\hspace*{-0.5cm} where {$\hat{w}$} is the direction vector to the
source in the LISA frame, {$\Delta \phi_m = \Omega L_m/2$, $k_{m \mp}
= 1 \mp \hat{w} \cdot \hat{n}$} appears in the accumulated phase due
to GW oscillation as the laser travels from one space-craft to another
(anti-)clockwise direction, {$\hat{a}_m$} is the normal vector from
LISA centre to each arm {$m$} and {$\xi_{m;+ \times}$} are the
responses of each arm to the two GW polarizations.

The SNR maximization for a particular direction is a constrained
optimization problem. The data combinations are obtained from the
eigen-vectors of the SNR squared matrix (averaged over the
polarizations) 
{\small
\begin{equation}
\rho^{(I)}_{(J)} = \left(f_{+}^{(I)} f_{+ (J)}^{*} + f_{\times}^{(I)} f_{\times (J)}^{*} \right)(\Omega)
\label{eigen}
\end{equation}
}
\vspace*{-0.35cm}

\hspace*{-0.6cm}
where {$f^{(I)}_{+,\times} = H_0 F^{(I)}_{+,\times}/n_{(I)}$},
{$H_0^2$} is the average signal energy over the GW polarization and
{$n_{(I)}^2$} is the noise PSD of {$Y^{(I)}$}. The eigen-values are
the instantaneous squared SNR for the optimal data streams. The 3
eigen vectors are given by:
%
\begin{eqnarray}
\vec{V}_+ &=& c_+ \f+ + c_\times \fx \,,\hspace{0.5cm} 
\vec{V}_\times = c_\times^* \f+ - c_+ \fx, \hspace{0.5cm} 
\vec{V}_0 = \f+^* \times \fx^* \,,
\end{eqnarray}
%
where the coefficient $c_\times = \fx^{*} \cdot \f+$ and
{\small
\begin{eqnarray}
c_+ = \frac{1}{2}\left[ |\f+ |^2 - |\fx |^2  + 
\sqrt{(|\f+ |^2 - |\fx |^2 )^2 + 4 |\f+ \cdot \fx^{*} |^2}\right] 
= {\rm snr}_+^2 - | \fx|^2 = -({\rm snr}_\times^2 - | \f+|^2) \,.
\end{eqnarray}
}
Note that the orthogonal pair of $\{\vec{V}_+,\vec{V}_\times\}$ is
obtained by the linear combination of $\f+$ and $\fx$ and hence lie in
$\{\f+,\fx\}$ --- polarization --- plane. The $2 \times 2$ matrix
which transforms $\{\f+,\fx\}$ to $\{\vec{V}_+,\vec{V}_\times\}$ is
traceless-hermitian.  Naturally, the direction orthogonal to this
plane contains no signal, i.e. a data stream obtained from $\vec{V}_0$
(orthogonal to $\{\f+,\fx \}$ plane) is a null stream.  Hence, the
triplet $\{\vec{V}_+,\vec{V}_\times, \vec{V}_0\}$ gives complete
directional information of the GW signal. The tracking coefficients for
the optimal data streams are
$\alpha_{(I)+,\times,0}=V_{+,\times,0}^{(I)*}/n_{(I)}$.  In summary,
the data stream (i) {$v_+ \equiv \alpha_{(I) +} Y^{(I)}$} gives maximum
directional SNR i. e. $\rm{snr_+}$, (ii) {$v_\times \equiv \alpha_{(I) \times}
Y^{(I)}$} gives the smallest non positive SNR i. e. ${\rm snr}_\times$
and (iii) $v_0 \equiv \alpha_{(I) 0} Y^{(I)}$} gives zero directional
SNR. At low frequency, $c_\times = 0$ and $v_+$ tracks the $+$
polarization while $v_\times$ tracks the $\times$ polarization of GW,
hence their subscripts \cite{NDPV03}.
\begin{figure}[!htb]
\begin{tabular}{cc}
(a) & (b) \\
\includegraphics[width=0.5\columnwidth, height =1.8in]{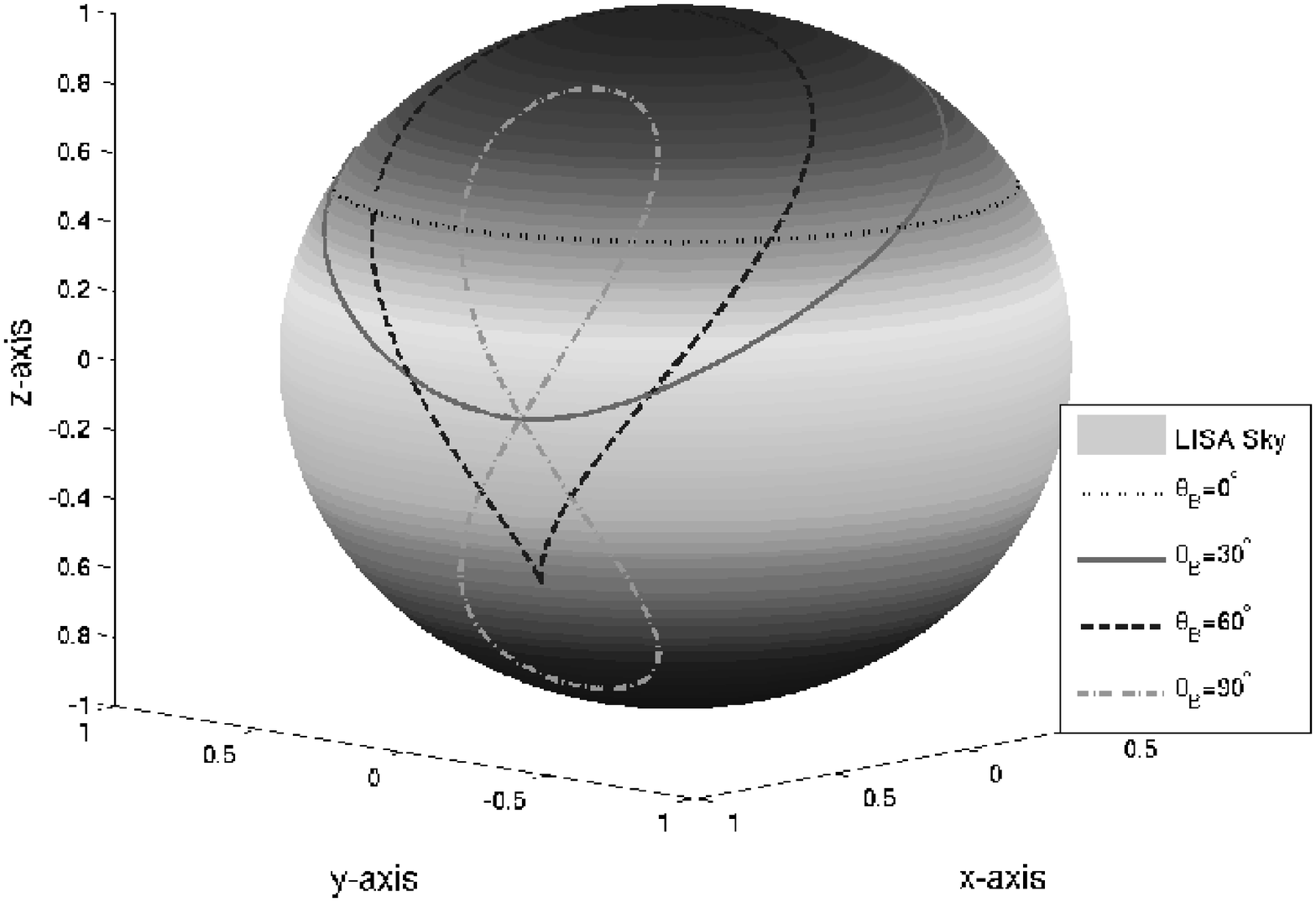}& \includegraphics[width=0.5\columnwidth,height=1.8in]{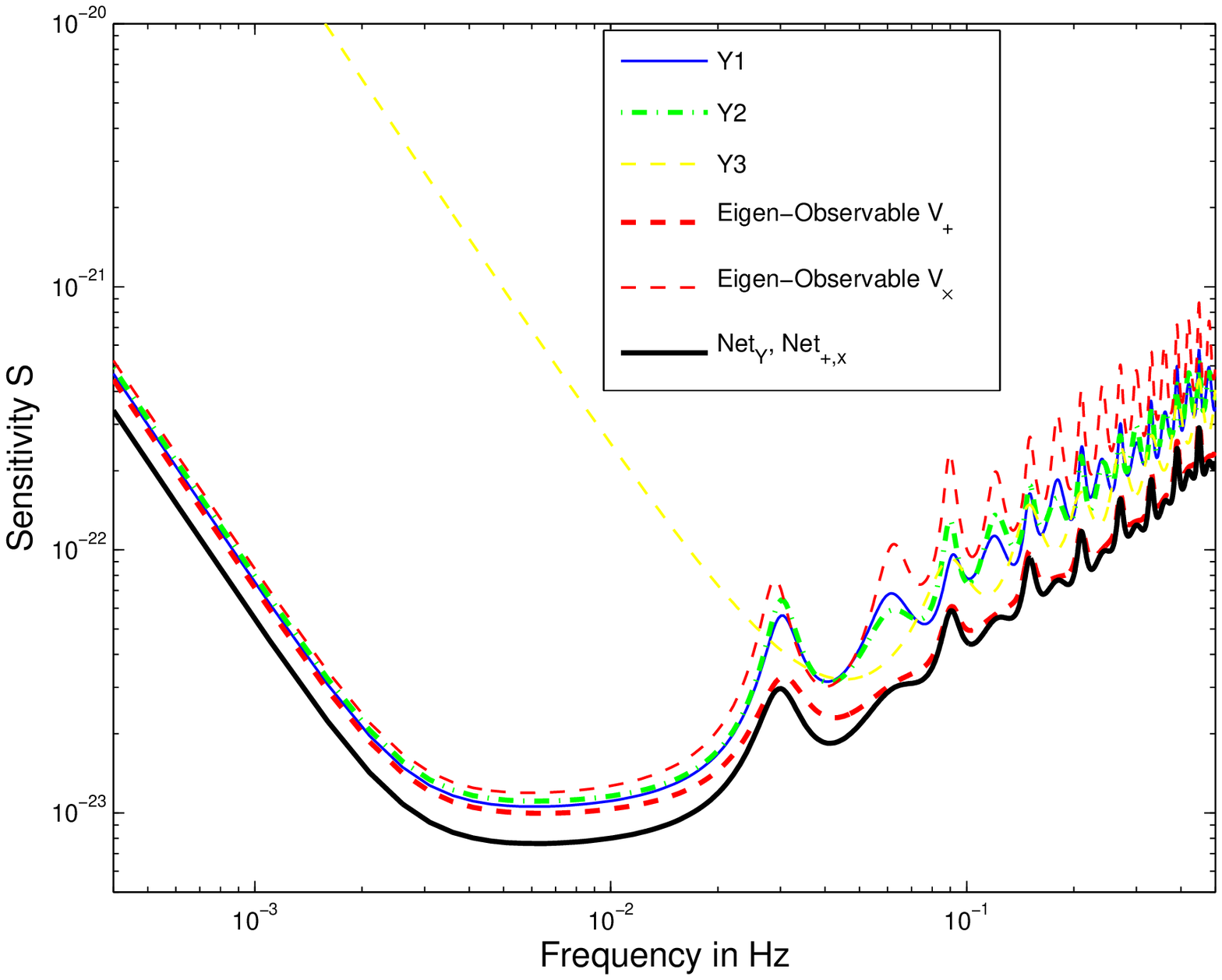}
\end{tabular}
\caption{(a) Source coordinates $\{X_L(t),Y_L(t),Z_L(t)\}$ in LISA frame for a fixed source at $\phi_B=200^\circ$ for various $\theta_B$. (b)
LISA Sensitivity for various data streams while tracking source at $\{\theta_B,\phi_B\}=\{60^\circ,200^\circ\}$ and $H_0=1$, $T=1$ year.}
\label{fig:sens}
\end{figure}

\subsection{Source tracking and Integrated SNR}
\label{sec:intSNR}
Due to LISA's motion, a fixed direction $\hat{w}_B =
\{\theta_B,\phi_B\}$ in barycentric frame appears to follow a track
{$\{\theta_L (t),\phi_L(t)\}$} in LISA's sky, see
Fig. \ref{fig:sens}(a). We track the direction and obtain the
integrated SNR as follows: We choose the optimal data
combination pertaining to the source direction {$\{\theta_L (t;
\hat{w}_B),\phi_L(t; \hat{w}_B)\}$}, at each time step, in LISA frame
(including Doppler shift correction to the frequency).  The
corresponding SNR at each time-step is referred to as instantaneous
SNR i.e. ${\rm snr_{+,\times}}$. We integrate the instantaneous SNR
as given below

\vspace*{-0.6cm}

{\small
\begin{eqnarray}
{\rm SNR}_{+, \times}^2 (\hat{w}_B) &=& \int_{0}^{T} {\rm snr}^2_{+, \times} (\theta_L (t; \hat{w}_B), \phi_L (t;\hat{w}_B )) dt \,.
\end{eqnarray}
}
\vspace*{-0.35cm} 

The network SNR while tracking $\hat{w}_B$ over 1 year period is
obtained by summing the squared integrated SNR's of the individual data streams
\footnote{Note that ${\rm SNR}_{\rm Net}^2(\hat{w}_B)$ can also be obtained by
summing the squared SNR's of the individual data streams
(trace of $\rho^{(I)}_{(J)}$) and integrating the resultant.} 
{\small
\be {\rm SNR}_{\rm
Net}^2(\hat{w}_B) = {\rm SNR}_+^2(\hat{w}_B)+{\rm
SNR}_\times^2(\hat{w}_B) = \sum_{I=1}^3 {\rm SNR}_{I}^2 (\hat{w}_B)\,.
\label{intsnr}
\ee
}
\vspace*{-0.3cm} 

In Fig.\ref{fig:sens}(b), we plot the LISA sensitivity $S=5/{{\rm
SNR}_{\rm int}}$ for a monochromatic source tracked with several data
combinations. ${\rm SNR}_{\rm int}$ is the integrated SNR along the
source track (in LISA frame) for a given combination.  For switching
combination, e.g. optimal streams $v_{+,\times}$, ${\rm SNR}_{\rm int}
= {\rm SNR}_{+,\times}$.

Fig.\ref{fig:sens}(b) displays the following features: (i) At low
frequencies ($f \le 3$ mHz), {$v_+$}, {$v_\times$} have similar
sensitivities and are proportional to $f^{-2}$ (similar to
$Y^{(1,2)}$); the data stream $Y^{(3)}$ ($= \zeta$) is insensitive to
GW. (ii) Above {$25$} mHz, all {$Y^{(I)}$}'s become comparable in
their sensitivities. The wavelength of GW $(\lambda_{\rm GW})$ is
comparable to $L$. This introduces geometry dependent features in the
LISA sensitivity curve. For example, $Y^{(3)}$ combination is most
sensitive when $f_{\rm gw} = f_L = 1/L= 60$ mHz or multiples of $f_L$.
As $v_+$ contains contribution from $Y^{(I)}$, this feature also
appears in its sensitivity, see Fig.\ref{fig:snrrat}(a).  The dips in
the integrated SNR of $v_+$ correspond to the zeros of $Y^{(3)}$ (at
$f= (2n-1)f_L/2$ mHz, $n$ is a positive integer).

\subsection{Beaming of Optimal Stream}

Although, tracking $\hat{w}_B$ with a network comprising of $Y^{(I)}$
--- Net$_Y$ --- and with a network of $v_+$ and $v_\times$ ---
Net$_{+,\times}$ --- gives the same integrated (as well as
instantaneous) SNR [see, for instance, Eq.(\ref{intsnr}) and
  Fig.\ref{fig:sens}(b)], we show in this section that they
possess completely different beaming properties.

%
\begin{figure}[htb!]
\includegraphics[width=0.75\columnwidth, height =2.in]
{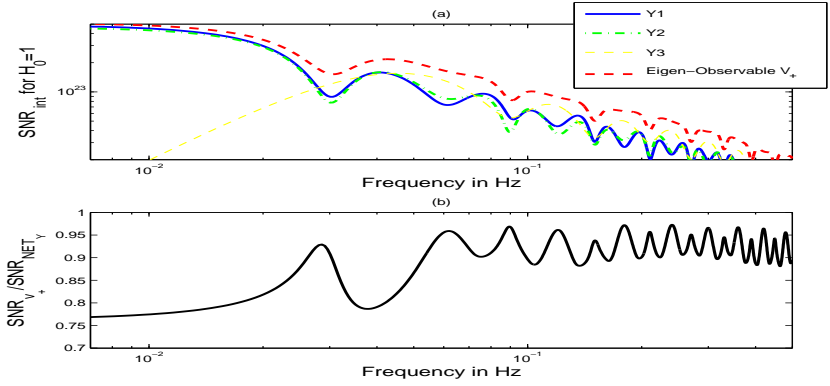}
\caption{(a) SNR of $v_+$ and $Y^{(I)}$ {\it vs} f, (b) ${\rm
SNR}+/{\rm SNR}_{\rm Net} {\it vs} f$ for tracking
$\{\theta_B,\phi_B\}=\{60^\circ,200^\circ\}$.}
\label{fig:snrrat}
\end{figure}

\begin{figure}[!htb]
\centering
\begin{tabular}{cc}
(a) & (b)\\
\includegraphics[width=0.5\columnwidth, height =2.5in]{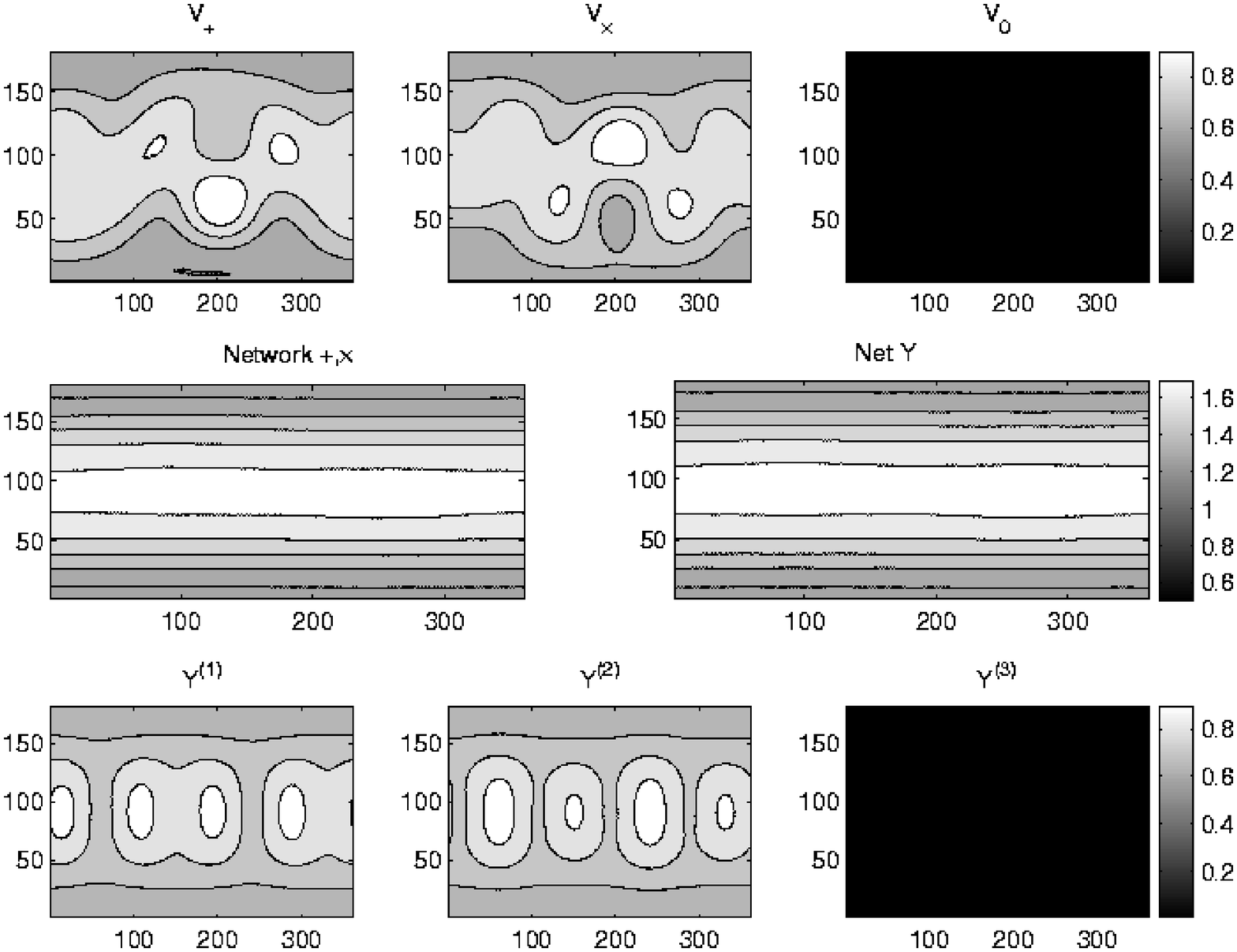}&
\includegraphics[width=0.5\columnwidth, height =2.5in]{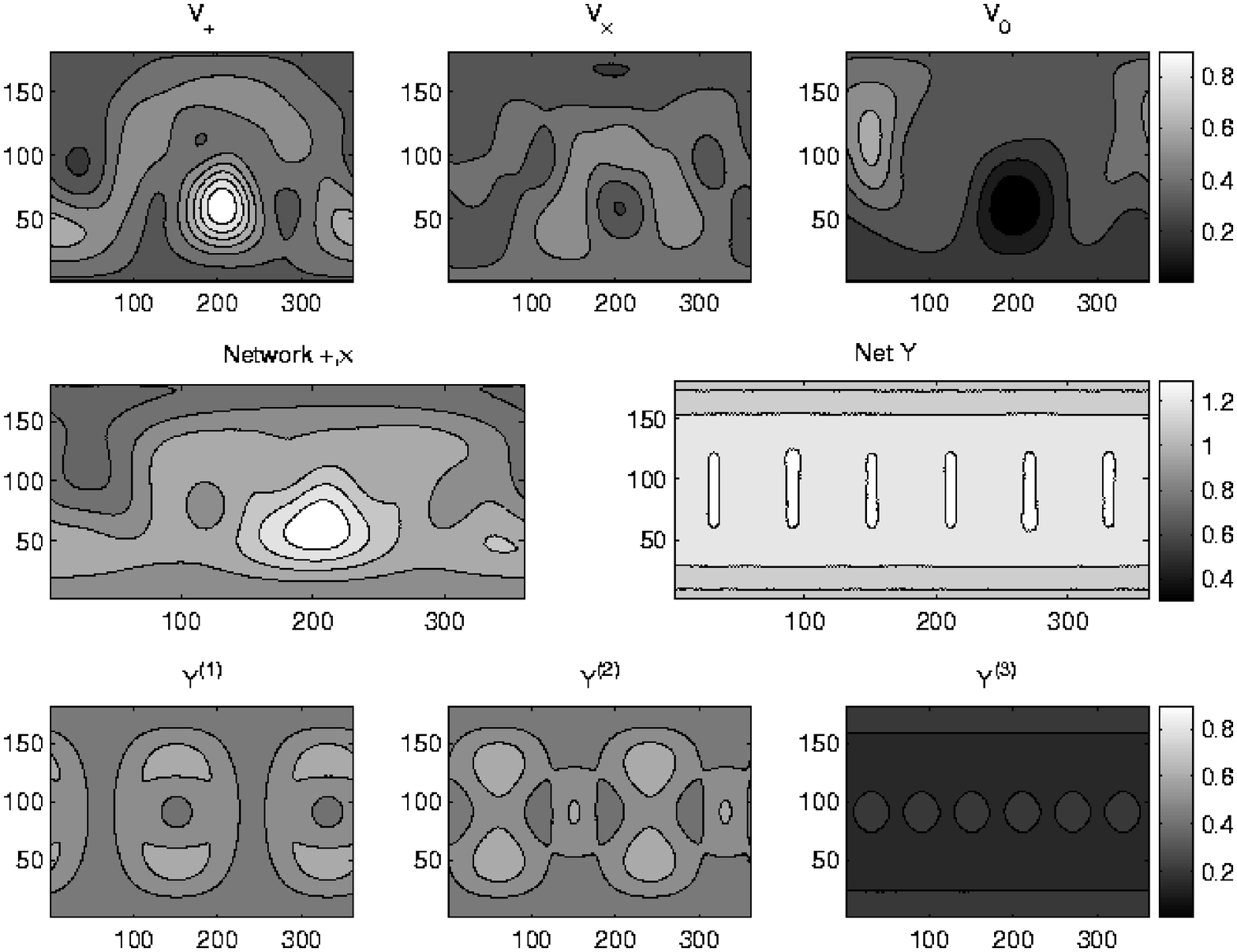}
\end{tabular}
\caption{Integrated squared SNR at (a) {$f=10$} mHz, (b) $f=25$ mHz and tracking $\{\theta_B,\phi_B\}=\{60^\circ,200^\circ\}$ with various data streams.
}.
\label{fig:ant_pat10}
\end{figure}

\begin{figure}[!htb]
\centering
\begin{tabular}{cc}
(a)&(b)\\
\includegraphics[width=0.5\columnwidth, height =2.5in]{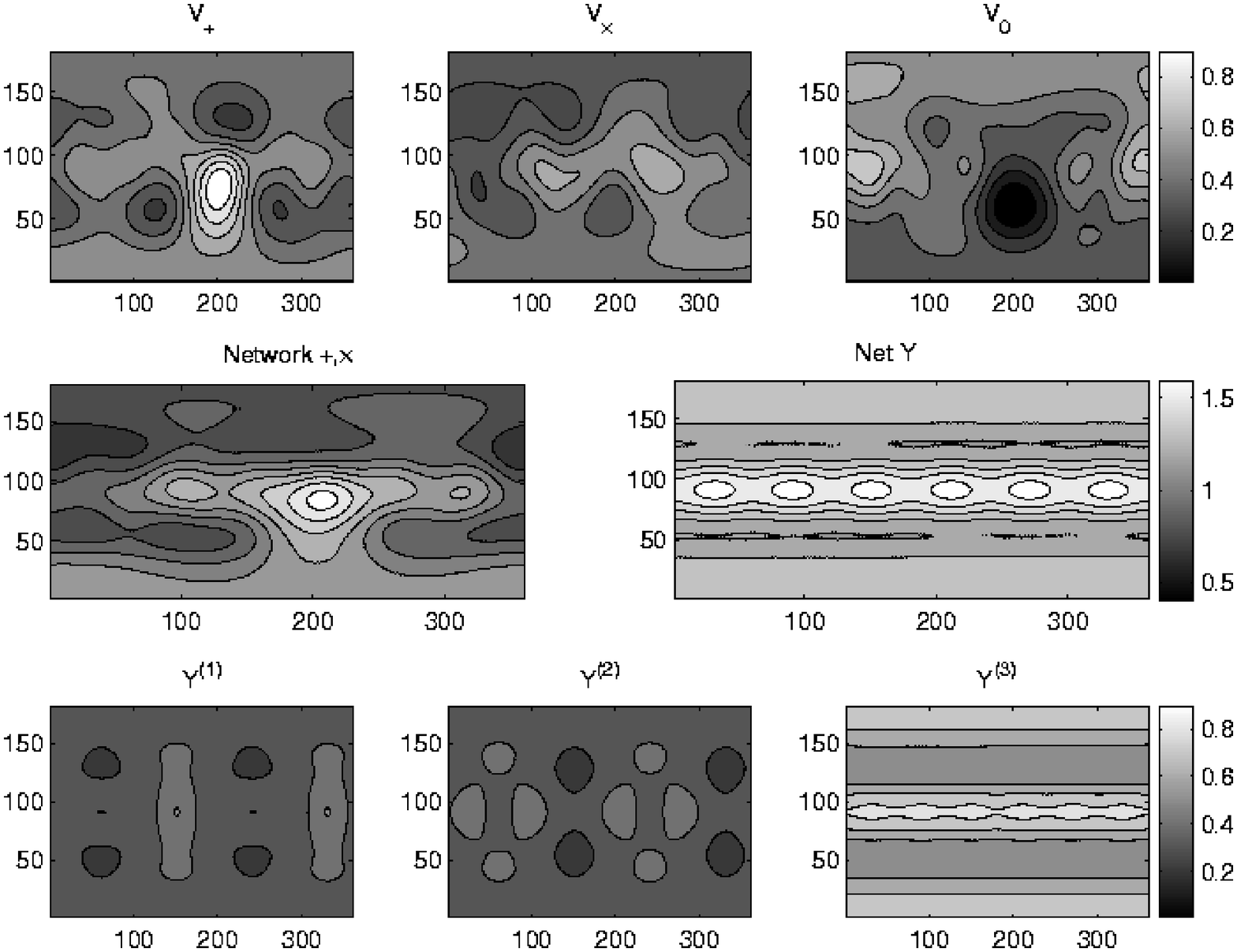}&
\includegraphics[width=0.5\columnwidth, height =2.5in]{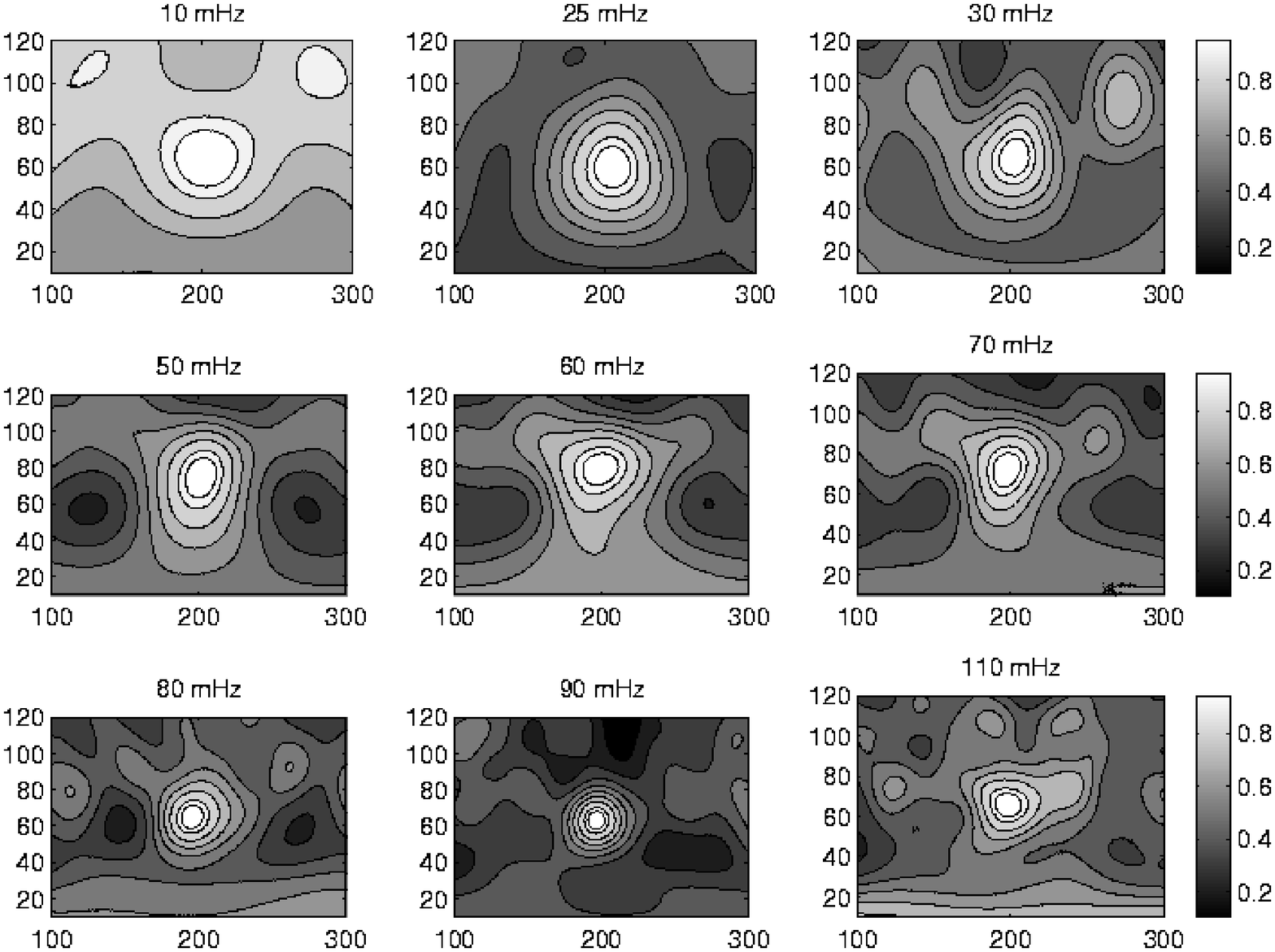}
\end{tabular}
\caption{Tracking $\{\theta_B,\phi_B\}=\{60^\circ,200^\circ\}$:
Integrated squared SNR (a) at {$f=50$} mHz for various data streams,
(b)for $v_+$ at different frequencies.
}.
\label{fig:ant_pat50}
\end{figure}

To demonstrate this, we divide the sky in the barycentric frame in 1
square degree patches. We observe each of this patch with the data
streams {$Y^{(I)}$} and integrate the squared SNR for a year along its
track in the LISA frame ($H_0=1$). In order to perform the same
exercise with {$v_+$}, {$v_\times$}, {$v_0$}, we choose to track an
arbitrary source direction
$\{\theta_B,\phi_B\}=\{60^\circ,200^\circ\}$. In
Fig. \ref{fig:ant_pat10},\ref{fig:ant_pat50}(a), contour plots of
integrated squared SNR are drawn (in barycentric frame) for all the
data streams.  The SNR is normalized with respect to ${\rm
SNR_+}$. The observations and the implications are as follows:\\
(i) The data streams $v_+$, $v_\times$ and $v_0$ are all beamed.  In
other words, the beam pattern of these streams peaks towards the
tracking direction, whereas the $Y^{(I)}$ streams are sensitive to a
large fraction of the sky; see, for instance,
Fig. \ref{fig:ant_pat10}(b), \ref{fig:ant_pat50}(a).  The contours of
the integrated SNR while tracking a particular direction give the
point spread function (psf) of the source.  In the language of the GW
data analysis literature, these contours can be related to the ambiguity
function in the source location parameter space. The width of the psf
determines the size of the template.\\
(ii) Net$_{+,\times}$ gives the same integrated SNR as that of
Net$_Y$. However, Net$_{+,\times}$ is highly beamed as opposed to
Net$_Y$.  The complimentary feature of Net$_{+,\times}$ and $v_0$ is
apparent from the bright and dark patch centered around the tracking
direction [see for instance, Fig. \ref{fig:ant_pat10}(b),
\ref{fig:ant_pat50}(a)].  This property can have a possible immediate
application in LISA data analysis.  In LISA, we expect to observe a
large number of GW sources from different sky locations in nearby
frequency bins. Thus, with the combination of Net$_{+,\times}$ and its
complementary null stream {$v_0$}, one could systematically suppress
other sky directions without compromising on SNR. \\ 
(iii) In the low frequency ({$f<10$} mHz), Net$_{+,\times}$ does not
show beaming and gives the same beam pattern as that of Net$_Y$.  This
is because the angular response of $Y^{(1),(2)}$ is that of a single
Michelson interferometer. The stream $Y^{(2)}$ differs from the
$Y^{(1)}$ by $45^\circ$ rotation [see
\cite{NDPV03},Fig.\ref{fig:ant_pat10} (a)] which makes the pattern
azimuthal invariant after tracking.  Further, the angular response of any TDI
combination will be limited by the signal antenna angular pattern
(size of earth's orbit $\sim \lambda_{\rm GW}$).\\
(iv) In Fig. \ref{fig:ant_pat10}(b) and \ref{fig:ant_pat50}(a), $v_+$
and Net$_{+,\times}$ are beamed along the tracking direction
$\{60^\circ,200^\circ\}$.
At $50$ mHz (being closer to $f\sim f_L$) , as opposed to at $25$ mHz, $Y^{(3)}$ is more sensitive
compared to $Y^{(1),(2)}$.


%



\begin{figure}[!htb]
\centering
\begin{tabular}{c}
\includegraphics[width=0.6\columnwidth, height =1.5in]{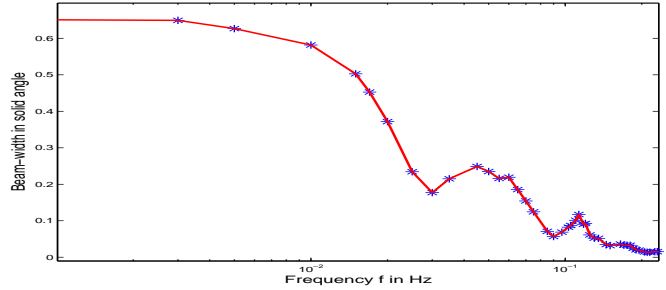}
\end{tabular}
\caption{Tracking $\{\theta_B,\phi_B\}=\{60^\circ,200^\circ\}$ with $v_+$:
beam-width (obtained at $90 \%$ level) {\it vs} $f$.}
\label{fig:sky_res}
\end{figure}

One would expect that beam-width of the optimal streams is a
monotonically decreasing function of frequency.  However, we find
super-imposed oscillations on the monotonically decreasing behavior
in the beam-width[see, Figs. \ref{fig:ant_pat50}(b),\ref{fig:sky_res}]\footnote{In
order to test this, we choose $v_+$ (maximum SNR) to track the source
at different frequencies. This is because, it is shown in
Fig.\ref{fig:snrrat}(b),, for {$f > 25$} mHz, {$v_+$} contributes more
than $80\%$ to Net$_{+,\times}$.}. Below, we summarise the features of
Fig. \ref{fig:sky_res}: At low frequencies ($<25$ mHz), the overall
beam-width of {$v_+$} decreases monotonically and is dominated by
{$Y^{(1,2)}$}.  At frequencies above $25$ mHz, one observes modulation
pattern in the beam-width (with frequency $f_L$).
More importantly, at $f \sim n f_L$, $Y^{(3)}$ contributes a large fraction
to SNR$_+$ as compared to the other two $Y$'s, see
Fig.\ref{fig:snrrat}(a). Hence, the beam-width of 
{$v_+$} is expected to carry the features of $Y^{(3)}$. 

As we know, {$Y^{(3)}$} is completely symmetric between the 3 arms. Its
angular response is insensitive in the neighborhood of LISA's zenith
as well as its equatorial plane whereas it is sensitive in the polar
window of $\theta_L:\{30^\circ,60^\circ\}$. A source at
$\theta_B=90^\circ$ follows a circular track in LISA sky
with $\theta_L=60^\circ$, see Fig.\ref{fig:sens}(a).
Hence, the source's track coincides with the
sensitive part of $Y^{(3)}$ beam-pattern which explains its maximum
integrated sensitivity along $\theta_B=90^\circ$.  As frequency
increases, especially near {$f=n f_L$}, due to the symmetry of
$Y^{(3)}$, for all {$\theta_L$}, the integrated antenna pattern of
$Y^{(3)}$ becomes invariant to azimuth $\phi_B$, see
Fig.\ref{fig:ant_pat10}(b),\ref{fig:ant_pat50}(a) which results in
increase in the beam-width.  This explains the sudden increase of
solid angle at {$n f_L$} in Fig.\ref{fig:sky_res}.

From Fig.\ref{fig:sky_res}, one can estimate the number of
non-overlapping patches (templates) required to cover the entire
sky. For instance, at $f \sim 10$ mHz, one requires $20$ such patches
whereas at $f=30$ mHz, this number goes to $60$.

\section{Conclusions}
\label{sec:IV}
In this work, we have studied the beaming properties of directionally
optimal data streams $v_+$,$v_\times$ and $v_0$ which contain the
complete information of the GW signal in a particular direction. We
have shown that they are beamed (after tracking the source for a year)
and could be useful in eliminating certain sky directions or for
consistency checks in LISA data analysis.

\section{Acknowledgement}
Thanks to Y. Chen, S.V. Dhurandhar, K.~R.~Nayak, J. Romano, L. Wen for
useful comments.  This work is supported by the Alexander von Humboldt
Foundation's Sofja Kovalevskaja Programme (Funded by the German
Ministry of Education and Research).

\end{document}